\documentclass[runningheads]{llncs}
\usepackage{graphicx}
\usepackage{algorithm}
\usepackage{algorithmic}
\algsetup{indent=2em}

\usepackage[colorlinks,citecolor=red,urlcolor=blue,bookmarks=false,hypertexnames=true]{hyperref} 

\begin{document}
\title{Local Search Trajectories over S-box Space\thanks{We want to acknowledge Stjepan Picek (\email{picek.stjepan@gmail.com}) for his suport.}}
%
%
\author{Ismel Mart\'{i}nez-D\'{i}az \inst{1} \and Carlos-Miguel Leg\'{o}n \inst{2} }
\authorrunning{Mart\'{i}nez-D\'{i}az \textit{et al.}}
%

\institute{ Ceibaco Lab, Cuba \\
	\email{ismel.martinez@nauta.cu} \and Havana University, Cuba \\
	\email{clegon58@gmail.com}}
\maketitle              
\begin{abstract}
The study of S-box properties relations is an interesting problem. In this work we develop and apply a local search method to create trajectories over S-box space. These trajectories shows the existence of an strong linear correlation between confusion coefficient variance, Transparency Order, Modified Transparency Order and Revised Transparency Order, under the Hamming Weight model. When the values of Confusion Coefficient Variance increases then the values of Transparency Order, the values of Modified Transparency Order beta zero, and the values of Revised Transparency Order beta zero, decreases, reflecting the same theoretical resistance against to Side-Channel Attacks by power consumption. As far as we know, it is the first time that Local Search trajectories are used to discover relations between cryptography properties.

\keywords{Local Search \and Confusion Coefficient Variance \and Transparency Order \and Modified Transparency Order \and Revised Transparency Order}
\end{abstract}
\section{Introduction}


Computer devices share information concerned to humans while they are interconnected: hosting or requesting data. This interchange of data is protected by symmetric cryptography \cite{caesar} and in particular by block ciphers, who has the function of ensure security and integrity of the data by encryption and decryption process using a key \cite{van2014encyclopedia,sehrawat2018lightweight}. 

In block ciphers design one of the main aspect is the substitution box definition (S-box): vectorial boolean functions that provide good confusion in the encryption (decryption) process \cite{avanzi2016salad}. This can be seen as two big problems:

\begin{itemize}
	\item How the S-box is defined.
	\item How the S-box ensure good resistance against an attack.
\end{itemize}

Concerning to the first problem, there are there main approaches:

\begin{enumerate}
	\item Algebraic construction.
	\item Heuristic search or generation -including  random generation-.
	\item Mixed methods: combination of 1. and 2.
\end{enumerate}

Algebraic construction was the first approach used to define S-boxes because the most known attacks in the literature: linear and differential attacks \cite{avanzi2016salad}, uses an algebraic point of view. The S-box of the Advanced Encryption Standard cipher (AES), is an example of an S-box obtained by algebraic construction \cite{nyberg1993differentially}. This S-box has a high Non-Linearity (NL) \cite{carlet2007nonlinearities}, one of the properties requires against linear attack \cite{avanzi2016salad}.

The second approach, heuristic search or generation, were included into both: design and cryptanalysis process, to make them more automatic. Initially the search methods were applied over the boolean functions \footnote{A boolean function \cite{van2014encyclopedia} can be considered as a particular kind of S-box with an input of $n$ bits and an output of 1 bit} space \cite{clark1998optimisation,clark2002metaheuristic}, with the same objective of increase the Non-Linearity value. But recently meta-heuristics has been used to obtain strong S-boxes, in particular, the evolutionary computation methods \cite{picek2015applications,vera2017algoritmos}.

The mix approach is the less study approach of the three. One example can be seen at \cite{de2018some}, where the mix method create some 4-bit cryptographic components by random generation and then apply an algebraic construction to obtain 8-bit S-boxes with high Non-Linearity.

Concerning to the second big problem, this work focuses on the theoretical resistance of an S-box against  Side-Channels Attacks by Power Consumption (SCA), the strongest attack presented in the actual literature. Some S-box properties measures this theoretical resistance and a fundamental line of research is to discover the relations between those properties.


This line of research is interesting and important because the properties intent to reflect the same resistance in different ways. The deductions of the properties starts from the SCA scheme with more or less generalization, sometimes under a particular leakage model of power consumption, sometimes using an specific statistical distinguisher.

One of the main properties to measure the theoretical resistance against to SCA was defined by Prouff in \cite{prouff2005dpa}, it is known as the Transparency Order (TO) and it appeared using the Hamming Distance model of power consumption and the difference of means as the statistical distinguisher. Chakraborty \textit{et al.} in \cite{chakraborty2017redefining} shows that the assumptions taking to deduce the Transparency Order were not ideal and the property only reflects the theoretical resistance under the Hamming Weight model of power consumption. In the same paper, the authors proposes some modifications and defines a new property: Modified Transparency Order (MTO), following a similar deduction to measure the resistance against to SCA under the Hamming Distance model. Lower MTO values represents better resistance.

Later in \cite{linotion}, Huizhong \textit{et al.} revisit the definitions of both: TO and MTO, providing a Revised Transparency Order definition (RTO) with a little conceptual difference with MTO. The difference remain in a mathematical aspect of statistical assumptions.

In a second line, Fei \textit{et al.} in~\cite{fei2012statistical} presents the Confusion Coefficient (CC) framework as a metric to measure the resistance against to Side-Channel Attacks, not only for this type of power consumption. This framework starts from a generalization and depends on two guessing key and a leakage function, measuring the probability that the leakages are different. Later, Picek \textit{et al.} in~\cite{picek2014confused} presents the Confusion Coefficient under the Hamming Weight model against SCA,  known as Confusion Coefficient Variance (CCV). Higher CCV values represents better resistance.


The notion of Transparency Order (Modified Transparency Order, Revised Transparency Order) and the notion of Confusion Coefficient Variance seems to share similarities. According to the different lines of properties deductions, it is difficult to realize a comparison analysis between the two approaches. Chakraborty \textit{et al.} in \cite{chakraborty2017redefining} refers that this will be a promising subject for research.


Lerman \textit{et al.} in~\cite{lerman2016comparing} makes a comparison between the Confusion Coefficient Variance and the Modified Transparency Order. The authors take into account both the Modified Transparency Order maximum value that measure the resistance under the Hamming Distance model and the Modified Transparency Order beta zero value (MTO-beta-zero) that measure the resistance under the Hamming Weight model. Only few S-boxes was used to realize the comparison and they concluded that the properties wasn't related at all.


In this work we show three new comparisons, the first one  between CCV and TO, the second one between CCV and MTO-beta-zero, and the third one between CCV and RTO-beta-zero. These comparisons goes in the form of a properties correlations.

For reveal the correlations we use the trajectories created by a Local Search method when it is applying over the S-box space. Local Search method is a meta-heuristic used to solve optimization problems. 
As we mention before, some meta-heuristics methods has been used to obtain S-boxes with high Confusion Coefficient Variance values or low Modified Transparency Order values, but those optimization process discard any S-boxes created in the intermediate steps, which could be used to realize data analysis at short or large scale.

In our results, as far as we know, is the first time that Local Search trajectories are used for an study of cryptography properties. Our goal is not to obtain S-boxes with good properties values but to offer a novel methodology for the cryptography research field.

\subsection{Preliminaries}

\subsubsection{Symmetric cryptography} The general scheme of a symmetric cryptography works as follow: Given a key $K$, a plain-text $X$, an encrypt function $E$ and a decryption function $D$; the cipher-text $X'$ can be created as $X' = E(X, K)$ and the plain-text will be recovered as $X = D(X', K)$.

S-boxes forms part of the encryption process $E$ and/or the decryption process D.

\begin{definition}

An NxM S-box is a vector boolean function  $F: \lbrace 0,1 \rbrace^{n} \rightarrow \lbrace 0,1 \rbrace^{m}$.

\end{definition}

\subsubsection{Side-Channel Analysis by Power Consumption} In the encryption (decryption) process on a computer device given $N$ known plain-texts (cipher-texts) $X(i)$, a power consumption is capture as a set of traces over time $T_{\dot{k}}(X(i))$. Those traces, and a hypothetical model of power consumption, are used to obtain every sub-key $\dot{k}$ of the secret key $K$. Both the traces and the leakages model exploit the evaluation of the S-box.

\subsubsection{Differential Power Attack} The main SCA attack is the DPA, which performs statistical analysis (calculate the difference of means) to retrieve the
secret sub-keys from the power consumption of cryptography devices. A \textit{single-bit} differential trace can be calculated by:

\begin{equation}
\label{dpa}
\Delta_{k, \dot{k}}(N,j) = \frac{\sum_{i = 1}^{N}V(X(i),k,j)T_{\dot{k}}(X(i))}{ \sum_{i = 1}^{N}V(X(i),k,j) } - \frac{\sum_{i = 1}^{N}(1 -V(X(i),k,j))T_{\dot{k}}(X(i))}{ \sum_{i = 1}^{N}(1 -V(X(i),k,j)) }
\end{equation}

Where in (\ref{dpa}), $k$ is a guessed sub-key and $V$ is a leakage binary function that depends on the known plain-text and the selected $j$-\textit{bit}.

\subsubsection{Power Leakages Models} The power leakages models more used in side-channels attacks by power consumption are: the Hamming Distance model and the Hamming Weight model~\cite{van2014encyclopedia}. The Hamming Distance model is interpreted as the result of the function $HW(\beta \oplus F(in \oplus k))$, where $\beta$ is a logic pre-charge of the cipher device, $in$ represents the input text and $k$ the sub-key used in the encryption process. The Hamming Weight function $HW(z), z \in \lbrace 0,1 \rbrace^{m}$, compute the number of ones in the boolean vector $z$ of $m$ components. In case of $\beta = \lbrace 0 \rbrace^{m}$, the model resulting of the function $HW(F(in \oplus k))$, it's renamed as the Hamming Weight model.

\subsubsection{Confusion Coefficient Variance} In \cite{fei2012statistical} is presented the Confusion Coefficient metric. This metric is computed for sub-keys $k_i$ and $k_j$ as:

\begin{equation}
\label{cc}
\kappa(k_i, k_j) = E[(W(k_i) - W(k_j))^2]
\end{equation}

Where in (\ref{cc}), $W$ represents the leakage function of the encryption process given an arbitrary input and the sub-key $k$.

Later, in \cite{picek2014confused} is proposed the Confusion Coefficient Variance (CCV) using the Confusion Coefficient (\ref{cc}) and the Hamming Weight model to simulate the leakages $W(k_i)$ and $W(k_j)$. It's formula, for all sub-keys $k_i, k_j, k_i \neq k_j$ and all input text $in$, is:

\begin{equation}
\label{eq:vcc}
CCV(F) = Var(E[(HW(F(in \oplus k_i)) - HW(F(in \oplus k_j)))^2])
\end{equation}

\subsubsection{Transparency Order} 
The Transparency Order (TO) is presented in \cite{prouff2005dpa}. This property try to catch the intrinsic S-box resistance against DPA attacks under the Hamming Distance model of power consumption.

Although the deficiencies finding in its formula deduction, TO still can be used to measure the theoretical resistance of an S-box against to DPA under the Hamming Weight model in a fast way \cite{diaz2017acelerando} applying the following formula:

\begin{equation}
\label{eq:to}
TO(F) =    m - \frac{1}{2^{2^{n}} - 2^{n}} \sum_{\alpha \in \lbrace 0,1 \rbrace^{n} - \lbrace 0 \rbrace^{n}}   {| n*2^{n} \sum_{x \in \lbrace 0,1 \rbrace^{n}}{HW(F(x) \oplus F(x \oplus \alpha))}| } 
\end{equation}

The TO property assume a DPA \textit{multi-bit} attack in the form of  $|\sum_{j} {  \Delta_{k, \dot{k}}(N,j)  }|$ (see \ref{dpa}).

\subsubsection{Modified Transparency Order} In \cite{chakraborty2017redefining}, the TO is modified because some deficiencies in its definition. The new property created is know as the Modified Transparency Order (MTO) and it's taking into account the cross-correlation spectrum of the components functions of the S-box $F=(F_1,...,F_m)$, denoted by $C_{F_i, F_j}(\alpha) = \sum_{ x \in \lbrace 0,1 \rbrace^{n}  }{(-1)^{  F_i(x) \oplus F_j(x \oplus \alpha) }}$. The property is computed as:

\begin{equation}
\label{eq:mto}
MTO(F) =  max_{ \beta \in \lbrace 0,1 \rbrace^{m}  } ( m - \frac{1}{2^{2^{n}} - 2^{n}} \sum_{\alpha \in \lbrace 0,1 \rbrace^{n} - \lbrace 0 \rbrace^{n}} \sum_{j = 1}^{m}   {| \sum_{i = 1}^{m} { (-1)^{\beta_{i} \oplus  \beta_{j} } C_{F_i, F_j}(\alpha) } | } )
\end{equation}

The Modified Transparency Order represents the theoretical resistance against to a Side-Channel Attack by Power Consumption under the Hamming Distance model. In particular, when $MTO(F)$ use only $\beta = \lbrace 0 \rbrace^{m}$ and discard all others $\beta$ values, it's denoted by (MTO-beta-zero) and the Hamming Distance model is reduced to the Hamming Weight model.

The MTO property assume a DPA \textit{multi-bit} attack in the form of  $\sum_{j} { | \Delta_{k, \dot{k}}(N,j) | }$ (see \ref{dpa}).

\subsection{Revised Transparency Order}

In the same fashion of MTO, the Revised Transparency Order represents the theoretical resistance against to Side-Channel Attack by Power Consumption under the Hamming Distance model and try to solve the TO deficiencies. But like TO, it assume the DPA \textit{multi-bit} in the form of  $|\sum_{j} { \Delta_{k, \dot{k}}(N,j) }|$(see \ref{dpa}). The property is computed as follow:

\begin{equation}
\label{eq:rto}
RTO(F) =  max_{ \beta \in \lbrace 0,1 \rbrace^{m}  } ( m - \frac{1}{2^{2^{n}} - 2^{n}} \sum_{\alpha \in \lbrace 0,1 \rbrace^{n} - \lbrace 0 \rbrace^{n}} | \sum_{j = 1}^{m}   { \sum_{i = 1}^{m} { (-1)^{\beta_{i} \oplus  \beta_{j} } C_{F_i, F_j}(\alpha) }  }| )
\end{equation}

When $RTO(F)$ use only $\beta = \lbrace 0 \rbrace^{m}$ and discard all others $\beta$ values, it's denoted by (RTO-beta-zero) and the Hamming Distance model is reduced to the Hamming Weight model.

\subsubsection{Local Search} Local Search is a meta-heuristic method. In the general scheme~\cite{delahaye2019simulated} the method starts from a random initial solution $s^{*}$, it finds neighborhood solutions $N(s^{*})$ and it moves from neighborhood to neighborhood while the objective function decreases (or increases in case of maximization). See Alg. \ref{lsgv}.

\begin{algorithm}

\caption{General scheme of Local Search}
\label{lsgv}

\begin{algorithmic}[1]

\ENSURE $s^{*}$ // Local optimum

\STATE $s^{*} \leftarrow random()$

\STATE search $\leftarrow$ \textbf{true}

\WHILE{search}
\STATE search $\leftarrow$ \textbf{false}

\FOR {$s_{i} \in N(s^{*})$} 
\IF {$f(s_{i}) < f(s^{*})$}
\STATE 			$s^{*} \leftarrow s_{i}$
\STATE search $\leftarrow$ \textbf{true}
\ENDIF
\ENDFOR
\ENDWHILE
\RETURN{$s^{*}$}
\end{algorithmic}

\end{algorithm}

The neighborhood construction depends of each problem. In this work we use the Hamming Weight leakages of S-boxes.

\section{Experiments and Results}

In this section we present a new Local Search method using Hamming Weight model. With this method we create trajectories over S-box space. The trajectories are used to detect the high degree of correlation between CCV and TO, CCV and MTO-beta-zero, and, CCV and RTO-beta-zero.

\subsection{Local Search using Hamming Weight Function}

We use the objective function $f = CCV(F)$ in a goal for increase the CCV value of the resulting S-boxes. We use either the next proposition (see Prop. \ref{proposition1}) that identify S-boxes with the same CCV value. 

\begin{proposition}
\label{proposition1}
Let $F_{A}$ and $F_{B}$ S-boxes defined in the same domain $\lbrace 0,1 \rbrace^{n}$ and image $\lbrace 0,1 \rbrace^{m}$. If $HW(F_{A}(x)) = HW(F_{B}(x)), \forall x \in \lbrace 0,1 \rbrace^{n}$ then $CCV(F_{A}) = CCV(F_{B})$.
\end{proposition}

\begin{proof}
The Hamming Weight of the S-boxes outputs will be equal for each entry $ x = in \oplus k \in \lbrace 0,1 \rbrace^{n}$, this imply that the expected value and the variance will be equal in the CCV formula (\ref{eq:vcc}) too.
\end{proof}

The new Local Search method has the following steps (see Alg \ref{nls}). In step 6 the method ensure the swapping of outputs with different Hamming Weight. Prop. \ref{proposition1} demonstrates that this condition on movement reduces the analysis of neighborhoods with the same CCV values. In the step 11 the climbing condition is checking. The method only stop when a local maximum is found.

\begin{algorithm}

\caption{Local Search method using Hamming Weight function (LS-HWF)}
\label{nls}

\begin{algorithmic}[1]

\ENSURE $F^{*}$

\STATE $F^{*} \leftarrow random()$

\STATE search $\leftarrow$ \textbf{true}

\WHILE{search}
\STATE search $\leftarrow$ \textbf{false}

\FOR {$\forall i,j \in \lbrace 0,1 \rbrace^{n}, j > i$}
\IF {$HW(F^{*}(i)) \neq HW(F^{*}(j))$}
\STATE $F \leftarrow F^{*}$
\STATE temp $\leftarrow F^{*}(i)$
\STATE $F(i) \leftarrow F^{*}(j)$
\STATE $F(j) \leftarrow$ temp
\IF {$CCV(F) > CCV(F^{*})$}
\STATE 			$F^{*} \leftarrow F$
\STATE search $\leftarrow$ \textbf{true}
\ENDIF
\ENDIF
\ENDFOR
\ENDWHILE
\RETURN{$F^{*}$}
\end{algorithmic}

\end{algorithm}

\subsection{Trajectories and correlations}

To study the correlations between CCV and TO, CCV and MTO-beta-zero, and, CCV and RTO-beta-zero, we design the next experiment:

\begin{enumerate}

\item Apply 30 runs of Alg. \ref{nls}.

\item For each run, in every $k$ climbing (steps 11-14 of Alg. \ref{nls}), create a sample of 30 random S-boxes with the same CCV of $F^*$ follow Prop. \ref{proposition1}. We remark that those S-boxes are not the result of the Local Search method and only are used for define the trajectories.

\item For each sample compute the mean of CCV and the mean of TO (MTO-beta-zero or RTO-beta-zero in other cases) of the 30 random S-boxes; and associate the pair $p_k(mean(CCV), mean(TO))$ with its correspondence $k$ climbing.

\item Finally, for each run, define its trajectory as the sequence $p_1, p_2, ..., p_q$ of the pairs associated to the total number of climbings $q$. Compute the linear correlation coefficient for the 30 trajectories and plot the trajectories on one same image.

\end{enumerate}

We execute the above experiment for the S-box spaces: 4x4, 5x5 and 8x8. We select those spaces in relation to the comparison making in~\cite{lerman2016comparing}, however we analyze more S-boxes. The number of S-boxes analyzed are limited by the Local Search in correspondence with the inherent stopping criteria of the method: reach a local optimum.

\subsubsection{CCV and TO correlations results}

Table \ref{ccvtoLinearcoefficient} shows, for every space, the mean and the standard deviation of the linear correlation coefficient of the 30 trajectories. As it can see, all means are negative and its absolute value are very high, and the standard deviation is very near to zero; all correlations values are neared to they respective mean. It reflects that, over all those S-box spaces, when the average of CCV values increases then the average of TO decreases almost lineally. Also we can check that, in order to increases the size of the space, the absolute value of the correlation mean increases too and the standard deviation decreases.

\begin{table}
\centering
\caption{Descriptive statistics on linear coefficient correlation values of the 30 Local Search trajectories obtained by LS-HWF for TO}
\label{ccvtoLinearcoefficient}

\begin{tabular}{|l|l|l|}
	\hline
	\textbf{S-box space} & \textbf{Mean} & \textbf{Standard deviation}  \\
	\hline
	4x4 &  -0.850364 &  0.121281 \\
	\hline
	5x5 &  -0.978528 &  0.011737 \\
	\hline
	8x8 &  -0.993467 &  0.002572 \\
	\hline
	
\end{tabular}

\end{table}

Table \ref{fig:table:correlations} Fig. a, b, c. shows, for every space, the 30 trajectories. It is clear that the relationship between the average values of these two variables (CCV and TO) can be adjusted approximately through a straight line. Instead 30 trajectories was created for every space, we can see that the trajectories are more close to the imaginary line that adjust them in correspondence of the higher size. We think that this behavior is related to the amount of the collected data; for 8x8 size, the trajectory is more large than the others because exists a higher range for CCV values and a higher climbing number.

\subsubsection{CCV and MTO-beta-zero correlations results}

We apply the same analysis for MTO-beta-zero. The results (see Table \ref{ccvMtoLinearcoefficient}) shows the same behavior with the little difference that the trajectories are more linear and the range of the MTO-beta-zero values are more sparse (see Table \ref{fig:table:correlations} Fig. d, e, f.), which helps to visualize the linear correlation in a better way.

\begin{table}
\centering
\caption{Descriptive statistics on linear coefficient correlation values of the 30 Local Search trajectories obtained by LS-HWF for MTO-beta-zero}
\label{ccvMtoLinearcoefficient}

\begin{tabular}{|l|l|l|}
	\hline
	\textbf{S-box space} & \textbf{Mean} & \textbf{Standard deviation}  \\
	\hline
	4x4 &  -0.978343 &  0.042503 \\
	\hline
	5x5 &  -0.997932 &  0.000837 \\
	\hline
	8x8 &  -0.988836 &  0.00449 \\
	\hline
	
\end{tabular}

\end{table}

\subsubsection{CCV and RTO-beta-zero correlations results}

For RTO-beta-zero analysis we create a sample of 1 S-box (which correspond to $F^{*}$) in every climbing. However, the statistical results (see Table \ref{ccvRtoLinearcoefficient}) shows the increasing linear correlation while the space grow. In contrast with MTO-beta-zero trajectories, only in 8x8 space the RTO-beta-zero trajectories has the expected behavior of strong correlation (see Table \ref{fig:table:correlations} Fig. g, h, i.).

\begin{table}
	\centering
	\caption{Descriptive statistics on linear coefficient correlation values of the 30 Local Search trajectories obtained by LS-HWF for RTO-beta-zero}
	\label{ccvRtoLinearcoefficient}
	
	\begin{tabular}{|l|l|l|}
		\hline
		\textbf{S-box space} & \textbf{Mean} & \textbf{Standard deviation}  \\
		\hline
		4x4 &  -0.730996 &  0.321642 \\
		\hline
		5x5 &  -0.937255 &  0.061624 \\
		\hline
		8x8 &  -0.982868 & 0.005616 \\
		\hline
		
	\end{tabular}
	
\end{table}

\subsubsection{General analysis}

The four metrics: CCV, TO, MTO-beta-zero, RTO-beta-zero, doesn't contradict each other when reflects the theoretical SCA resistance under the Hamming Weight leakage model. The higher CCV values is an indicator of lower TO values, lower MTO-beta-zero values and lower RTO-beta-zero values.

TO trajectories and MTO-beta-zero trajectories has very similar behavior for every correspondence space. But RTO-beta-zero trajectories only shows a more strong linear correlation for the 8x8 S-box space.

We consider that the CCV is a good theoretical metric for S-box design to reflect the resistance against to SCA under the Hamming Weight model. Still, we can't say anything about which one of the four metrics is the fastest to do it, or how good is the MTO (RTO) to reflect the resistance against to SCA under the Hamming Distance model.

\begin{table}
	\centering
	\caption{ \textbf{Figures} (a) CCV vs TO, 4x4 S-box space. (b) CCV vs TO, 5x5 S-box space. (c) CCV vs TO, 8x8 S-box space. (d) CCV vs MTO-beta-zero, 4x4 S-box space. (e) CCV vs MTO-beta-zero, 5x5 S-box space. (f) CCV vs MTO-beta-zero, 8x8 S-box space. (g) CCV vs RTO-beta-zero, 4x4 S-box space. (h) CCV vs RTO-beta-zero, 5x5 S-box space. (i) CCV vs RTO-beta-zero, 8x8 S-box space.}
	\label{fig:table:correlations}
	
	\begin{tabular}{cc}

		\includegraphics[width=17em]{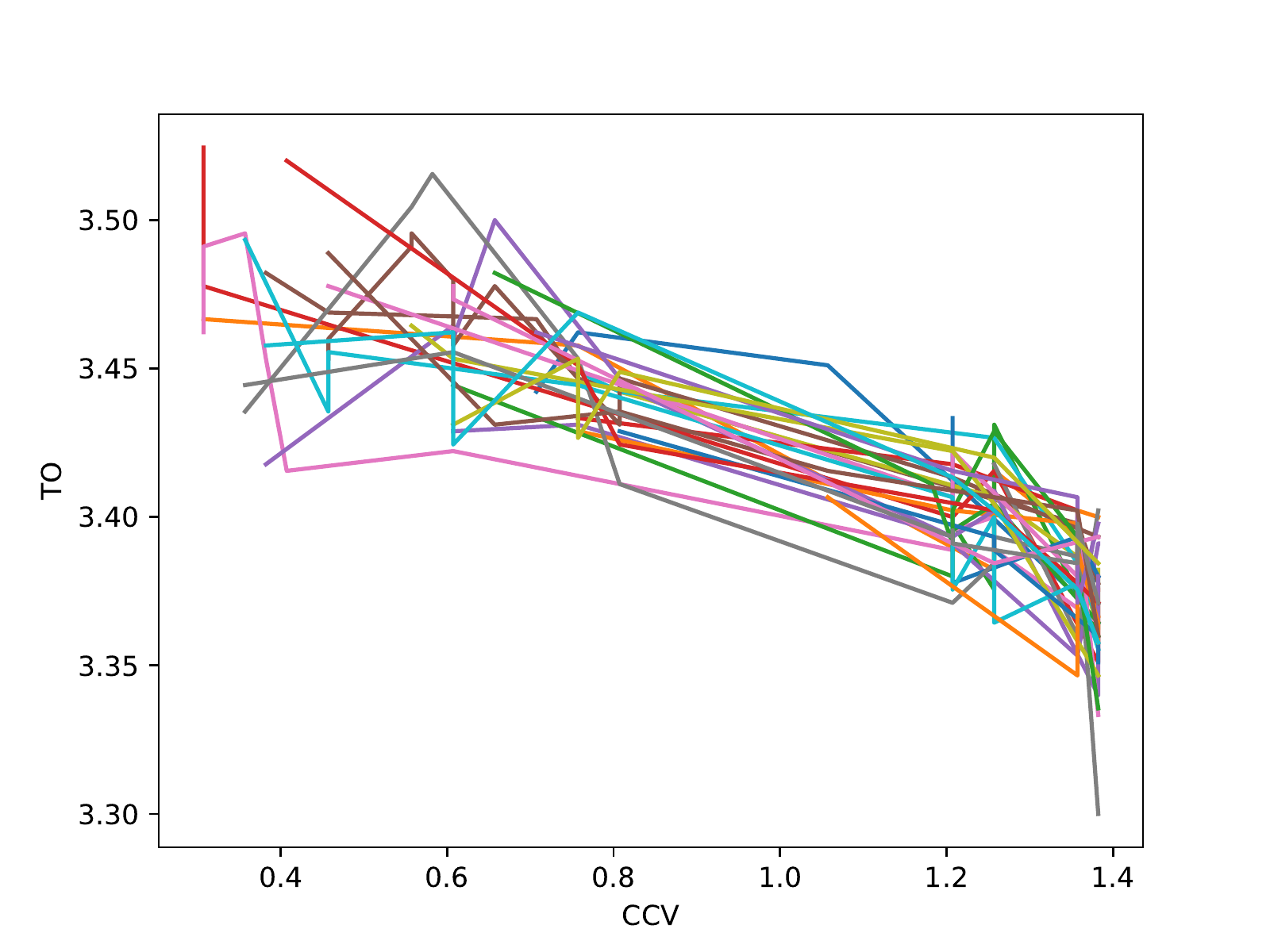}
		(a)

		&

		\includegraphics[width=17em]{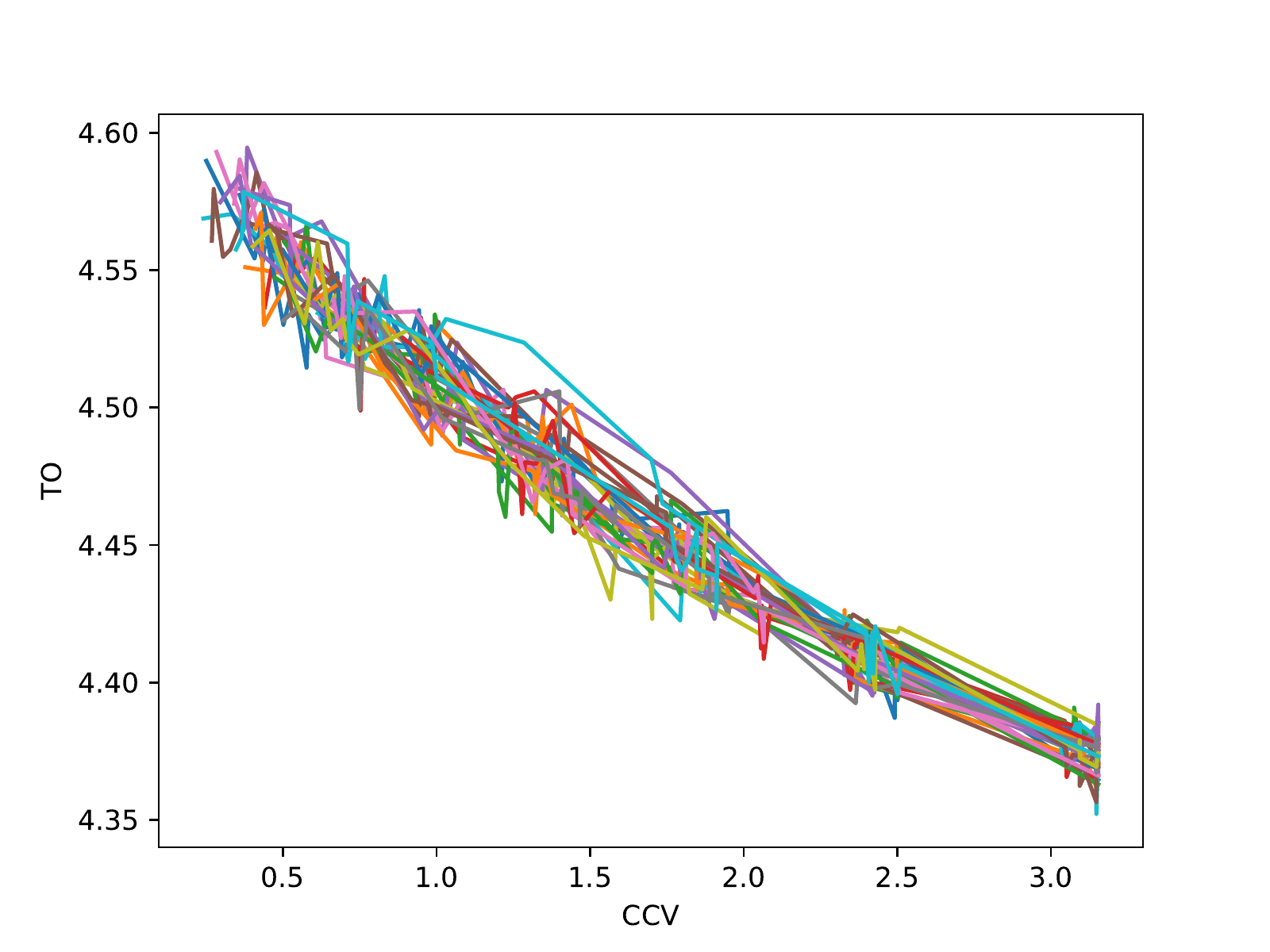}
		(b)
		\\

		\includegraphics[width=17em]{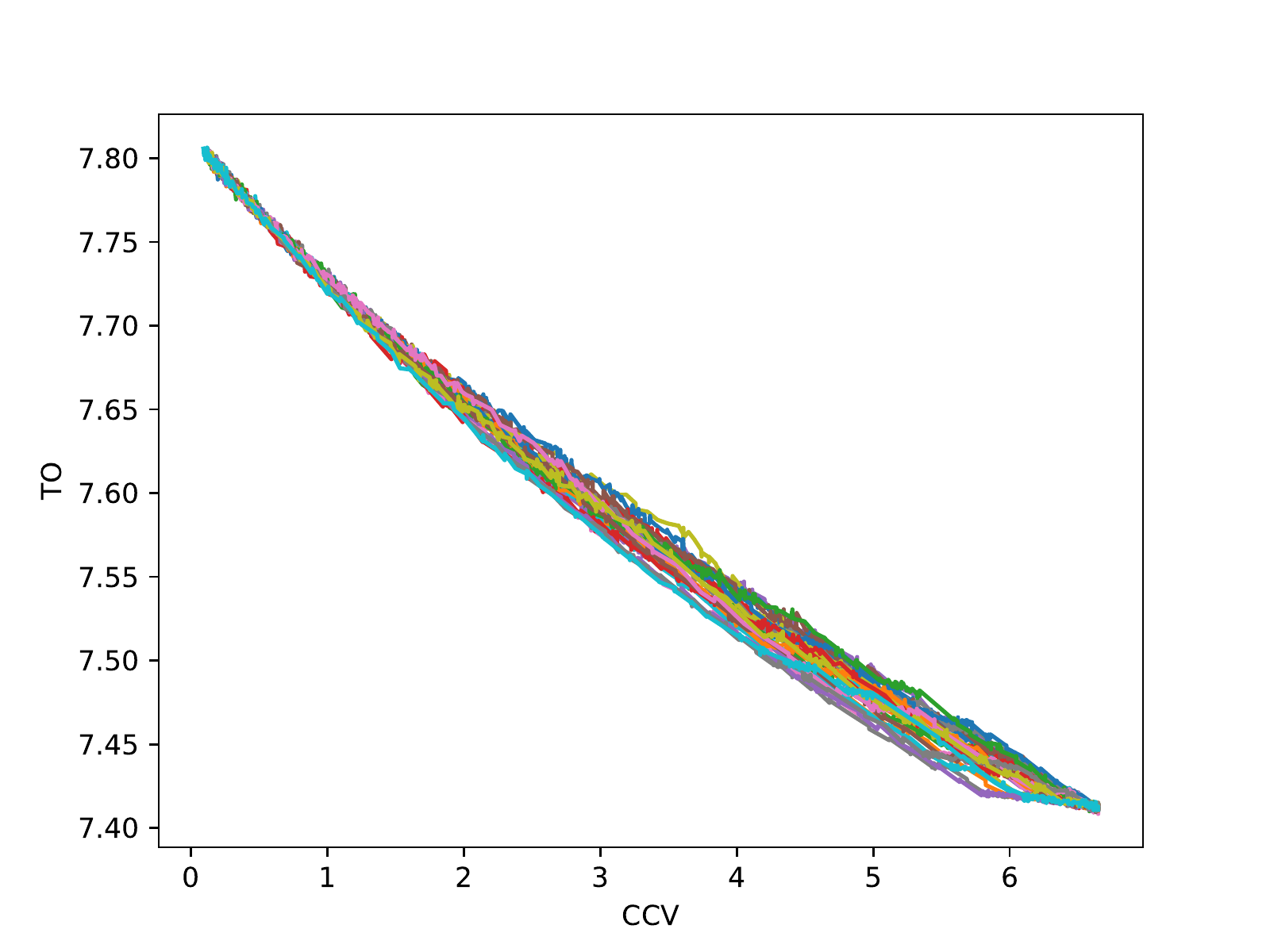}
		(c)
		&

		\includegraphics[width=17em]{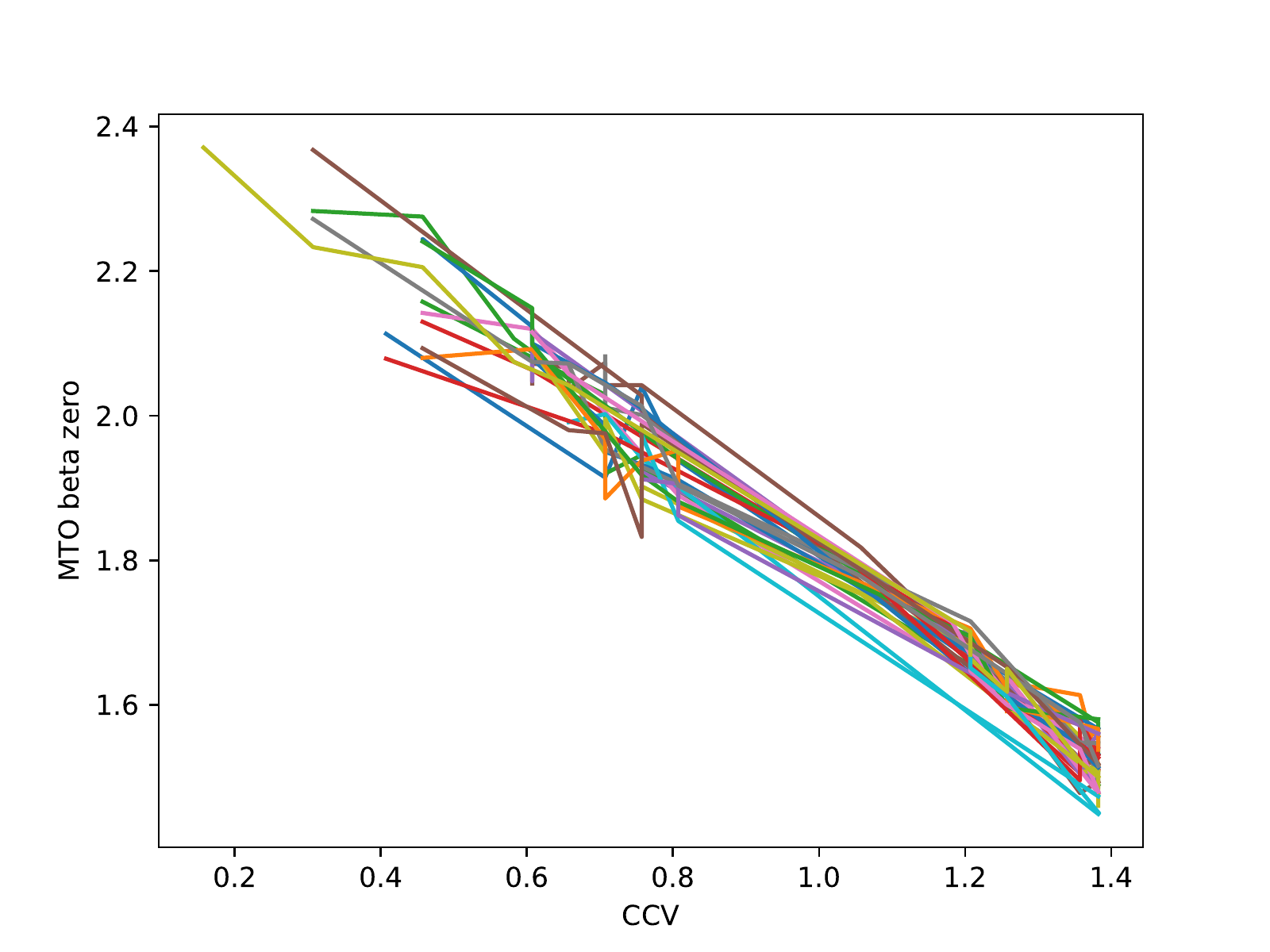}
		(d)
		\\

		\includegraphics[width=17em]{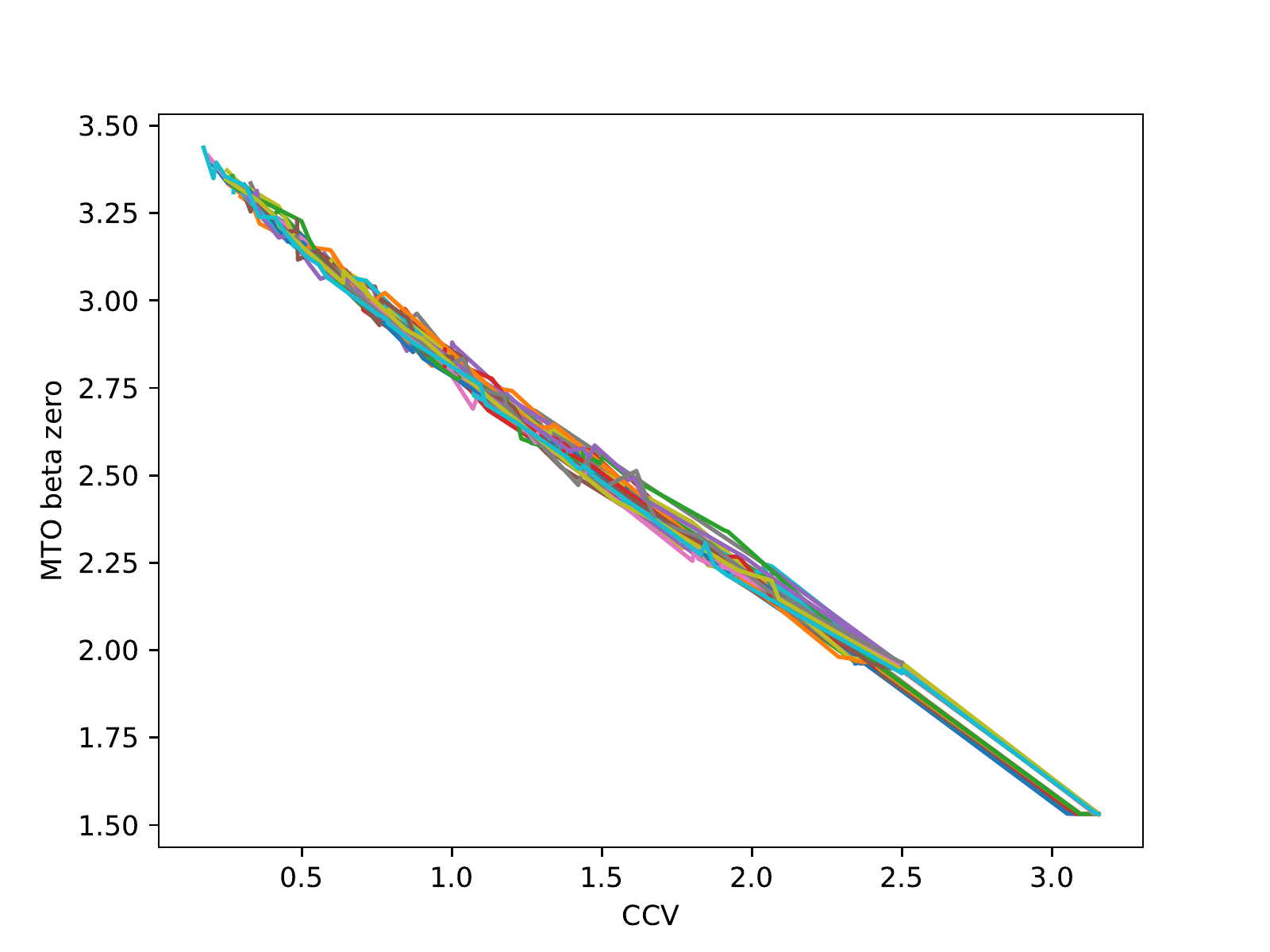}
		(e)
		&

		\includegraphics[width=17em]{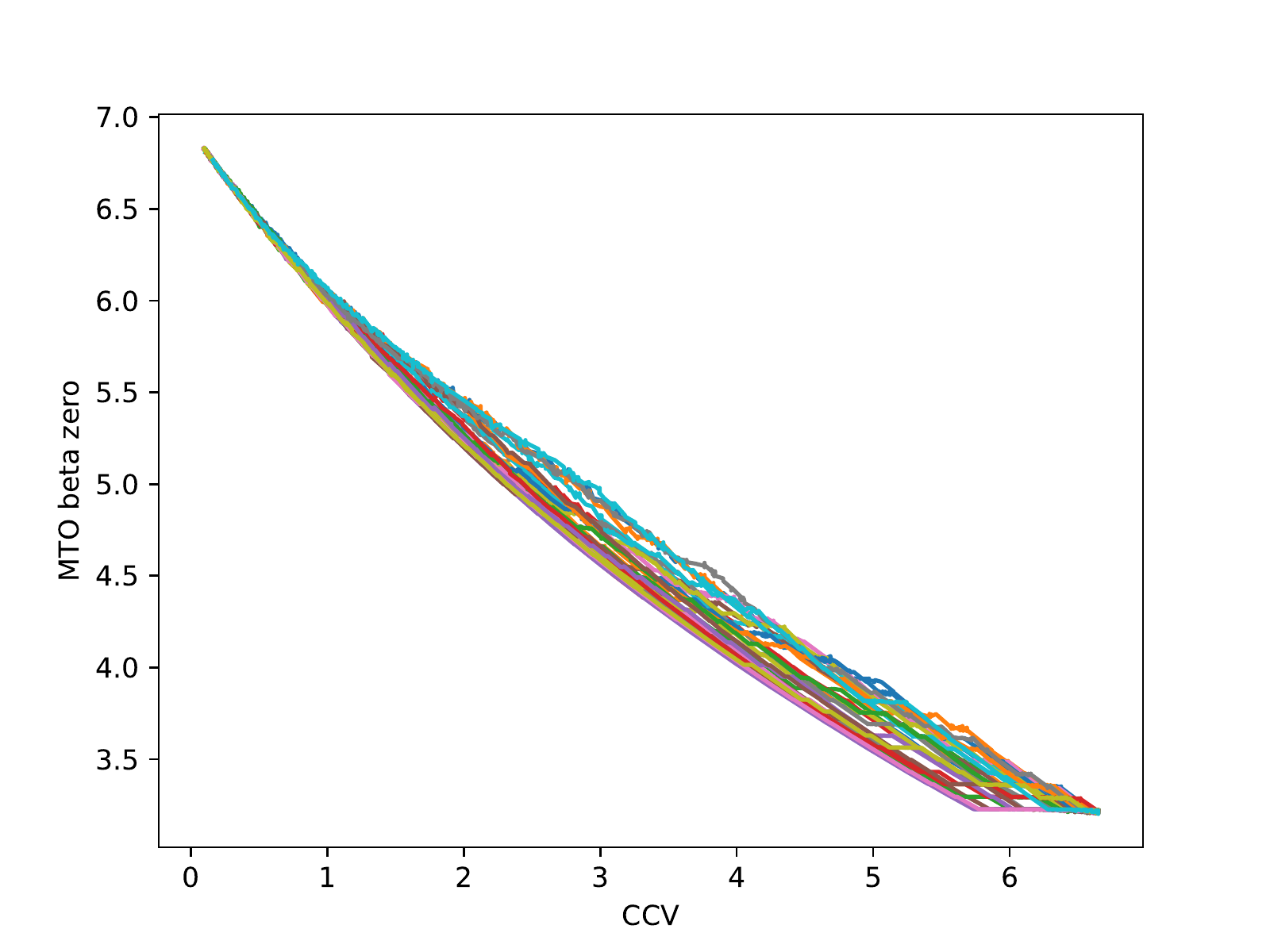}
		(f)
		\\

		\includegraphics[width=17em]{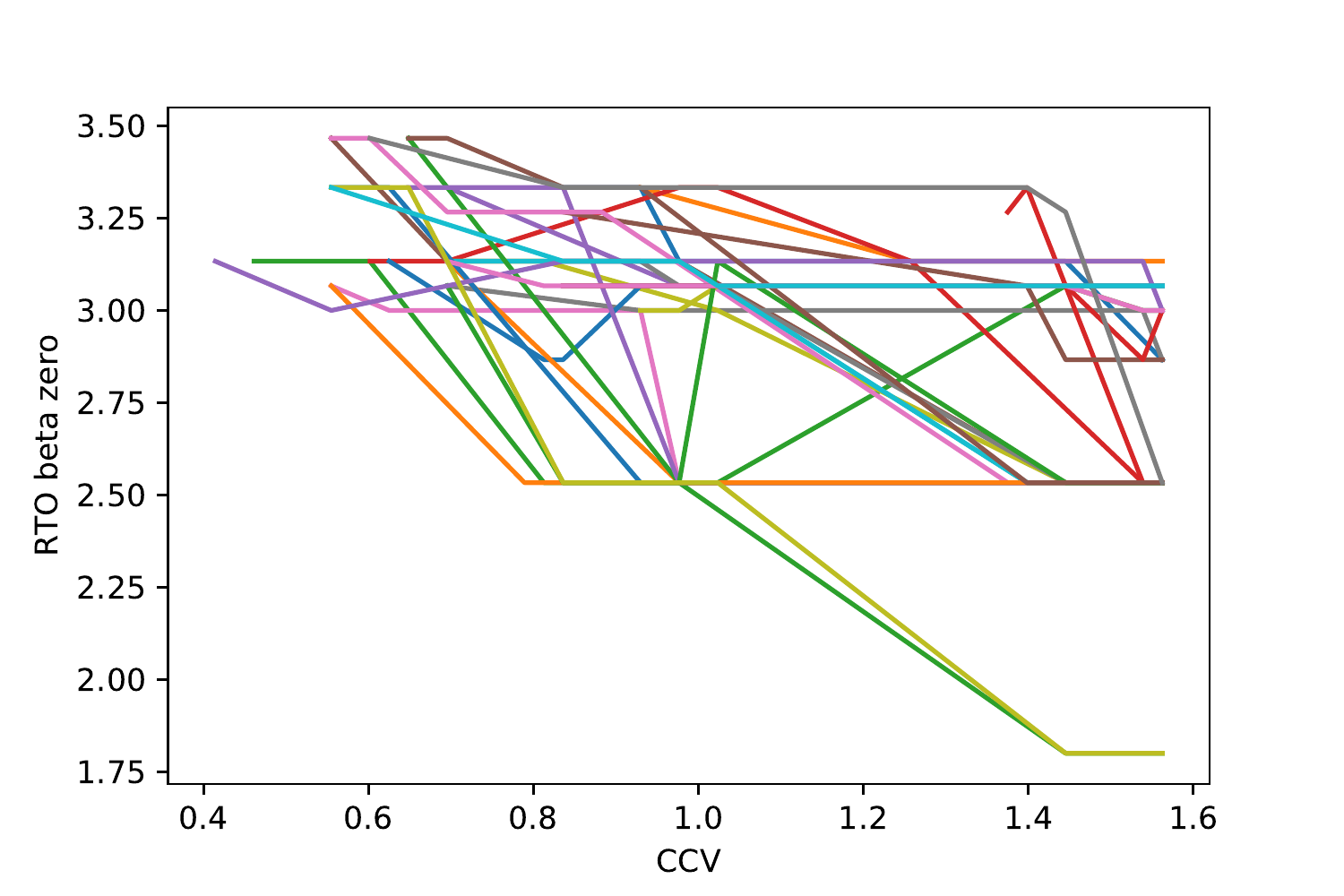}
		(g)
		
		&
		\includegraphics[width=17em]{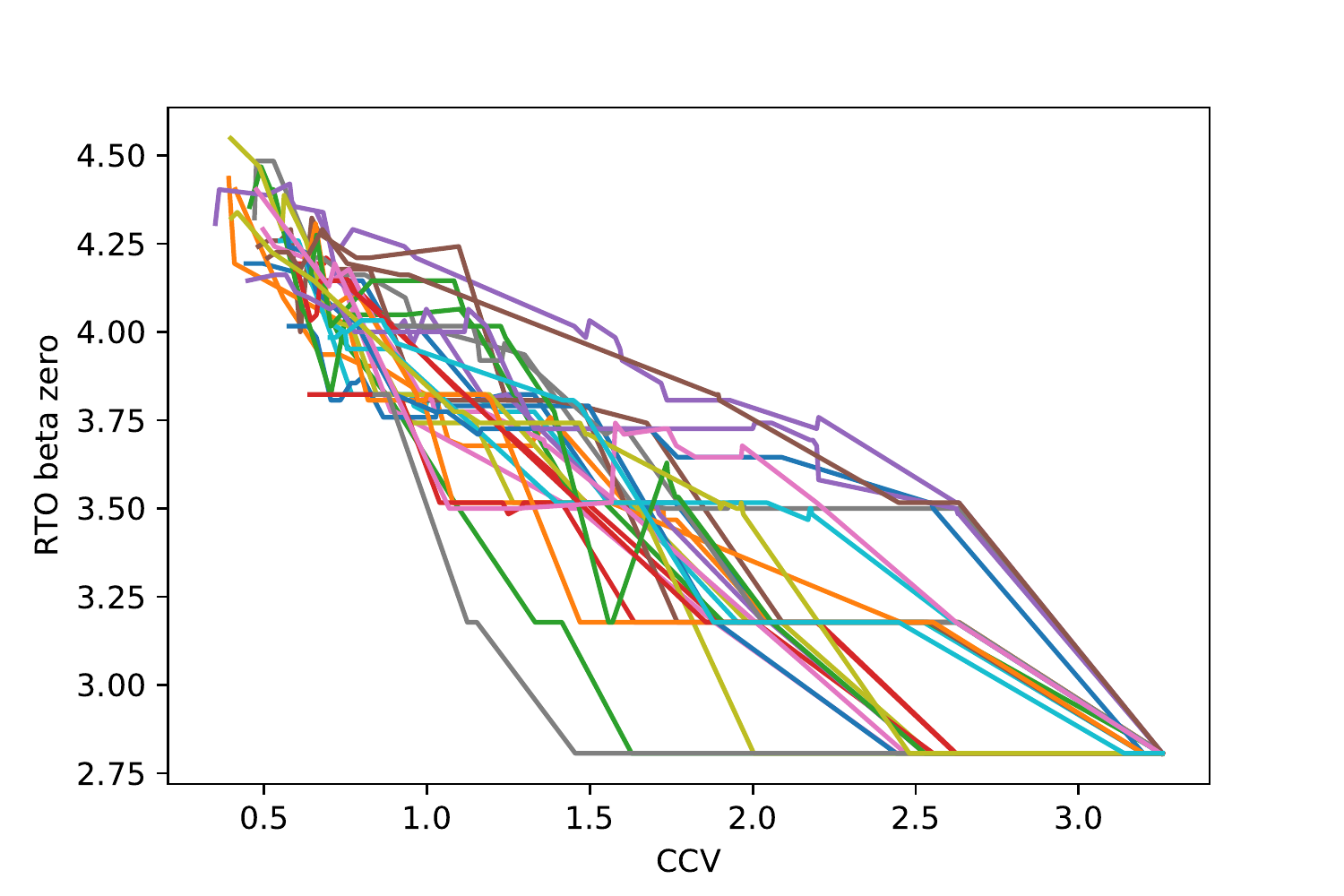}
		(h)
		
		\\
		
		\includegraphics[width=17em]{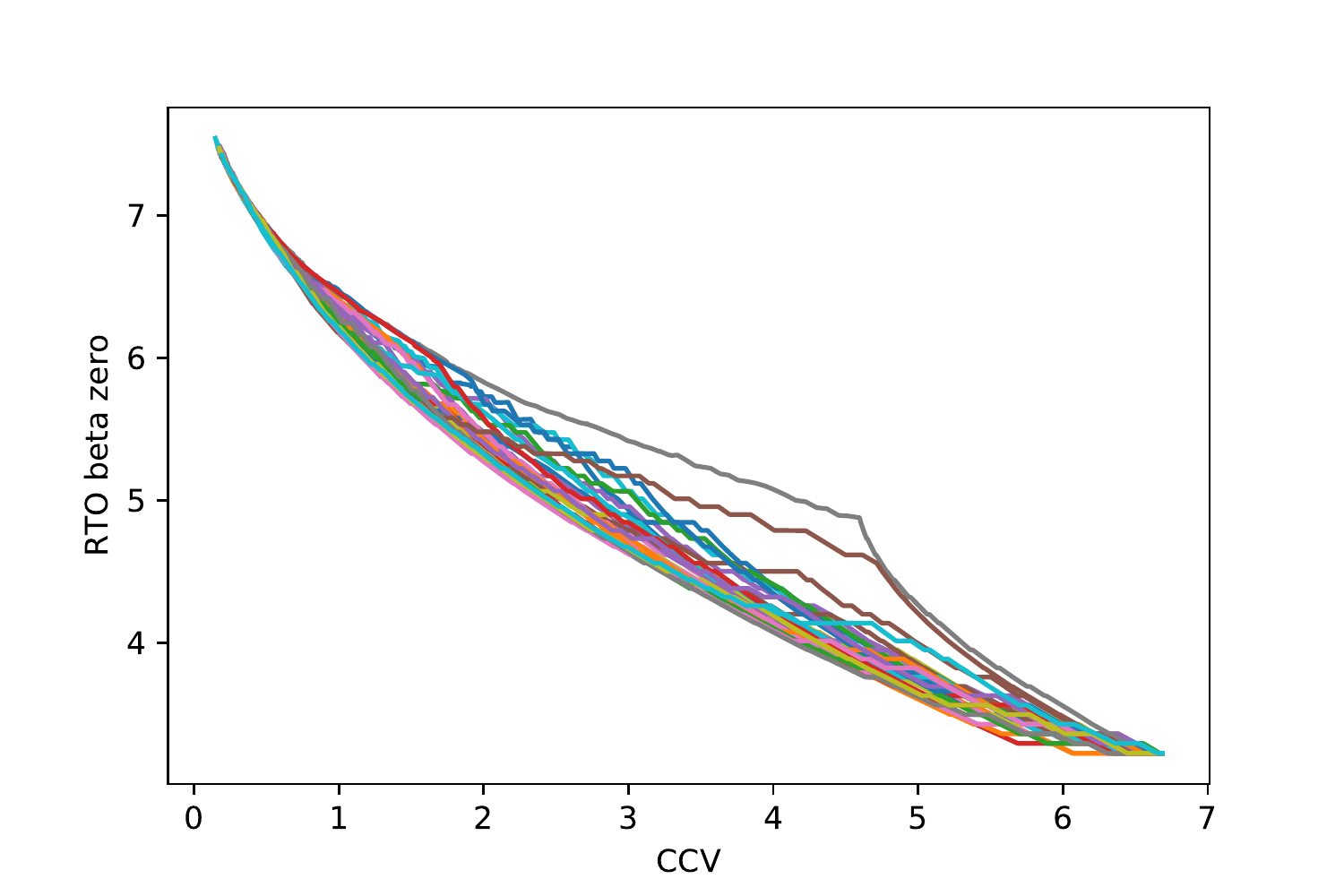}
		(i)
		
		&
		\\

	\end{tabular}
	
\end{table}

\newpage

\section*{Conclusions}

The Local Search trajectories over S-box space is a good methodology for study the correlations of S-box properties. An application of this methodology is the resulting almost linear correlation between Confusion Coefficient Variance and the Transparency Order, the Modified Transparency Order, and the Revised Transparency Order, under the Hamming Weight leakage model.

Some future perspectives of this work will be to discover the relationship between the Confusion Coefficient Variance property and the Non-Linearity property. Also to follow a new definition of CCV but under the Hamming Distance model, for make a comparison with the Modified Transparency Order.


%
%
%
\bibliographystyle{splncs04}
\bibliography{sample}

\end{document}